\documentclass[rnote]{aa} % for the research notes
\usepackage{graphicx}
\usepackage{txfonts}
\usepackage{natbib}
\usepackage{longtable}
\usepackage{lscape}
\usepackage{appendix}
\begin{document}

\title{Luminosities of Carbon-rich Asymptotic Giant Branch stars in the Milky Way}

\subtitle{}

\author{
  R. Guandalini\inst{1,2}
\and
  S. Cristallo\inst{3}
}

\institute{
  University of Hertfordshire, Physics Astronomy and Mathematics, Hatfield, AL10 9AB, UK\\
  \email{roald.guandalini@gmail.com}
\and
  Dipartimento di Fisica, Univ. di Perugia, via Pascoli, 06123 Perugia, Italy\\
\and
  Osservatorio Astronomico di Teramo (INAF), via Maggini snc 64100 Teramo, Italy\\
  \email{cristallo@oa-teramo.inaf.it}
}

\date{Received ; Accepted }

% \abstract{}{}{}{}{}
% 5 {} token are mandatory

\abstract
{Stars evolving along the Asymptotic Giant Branch can become
Carbon-rich in the final part of their evolution. They
replenish the inter-stellar medium with nuclear processed material
via strong radiative stellar winds. The determination of the
luminosity function of these stars, even if far from being
conclusive, is extremely important to test the
reliability of theoretical models. In particular, strong
constraints on the mixing treatment and the mass-loss rate can be
derived.}
{We present an updated Luminosity Function of Galactic Carbon
Stars obtained from a re-analysis of available data already
published in previous papers.}
{Starting from available near- and mid-infrared photometric data,
we re-determine the selection criteria. Moreover, we take
advantage from updated distance estimates and Period-Luminosity
relations and we adopt a new formulation for the computation of
Bolometric Corrections. This leads us to collect an improved
sample of carbon-rich sources from which we construct an updated
Luminosity Function.}
{The Luminosity Function of Galactic Carbon Stars peaks at
magnitudes around $-$4.9, confirming the results obtained in a
previous work. Nevertheless, the Luminosity Function presents two
symmetrical tails instead of the larger high luminosity tail
characterizing the former Luminosity Function.}
{The derived Luminosity Function of Galactic Carbon Stars
matches the indications coming from recent theoretical
evolutionary Asymptotic Giant Branch models, thus confirming the
validity of the choices of mixing treatment and mass-loss history.
Moreover, we compare our new Luminosity Function with its
counterpart in the Large Magellanic Cloud finding that the two
distributions are very similar for dust-enshrouded sources, as expected from stellar
evolutionary models. Finally, we derive a new fitting formula aimed
to better determine Bolometric Corrections for C-stars.}


  \maketitle
%
%________________________________________________________________

\section{Introduction}

After the exhaustion of their central helium, stars with initial
masses between 0.8 and 8 M$_\odot$ evolve through the Asymptotic
Giant Branch phase (hereafter AGB). These stars efficiently
pollute the Inter-Stellar Medium (hereafter ISM) by ejecting their
cool and expanded envelopes via strong stellar winds powered by
radial pulsations and radiation pressure on dust grains. Moreover,
being extremely luminous objects, they provide the dominant
contribution to the integrated light of aged stellar populations.
A detailed modeling of their theoretical evolution and an
extensive study of their observational properties is therefore
mandatory.\\

During the AGB phase, nuclear processed material from stellar
interiors can be carried to the surface by means of the so called
Third Dredge Up episodes (see \citealt{stra06} and references
therein). A large quantity of $^{12}$C, synthesized via the
3$\alpha$ reaction, can be mixed into the envelope, thus
increasing the surface C/O ratio and making this object a carbon
star. There are still many open problems affecting our knowledge
of the physics of AGB stars. Among major theoretical
uncertainties, we highlight the treatment of the mixing
and the mass loss history. Both can alter the surface C/O
in models and the time spent by the star in the C-rich
phase. From the observational point of view, major problems in
determining the Luminosity Function of Carbon Stars are a proper
evaluation of the Infrared (IR) contribution (in particular for
Mira variables) and distance estimates. In the past, this
led to the formulation of the so-called "carbon stars
mystery" \citep{iben} and to a long disagreement
between observers and theoretical modelers (see e.g. \citealt{izza}).\\
In this paper we present a new estimate for the observational
Luminosity Function derived from a sample of Galactic Carbon Stars
(Luminosity Function of Galactic Carbon Stars, hereafter LFGCS).
This quantity links observed quantities (the
luminosity and the surface C/O ratio) with theoretical
properties of stellar models (in particular
their core masses, which depend on the treatment of mixing
and the adopted mass-loss law: see \citealt{cris11} and
references therein). A precise and unbiased observational LFGCS
constitutes therefore a fundamental yardstick to check the
reliability of stellar theoretical models and, in addition, Galactic
chemical evolution models. This is not an easy task at all, mainly
hampered by the difficulties in measuring the distances of these
dust-enshrouded objects. \\

\citet{gua06} derived the observational LFGCS, stressing that
meaningful C-stars luminosities cannot be extracted from fluxes
obtained at optical and near-IR wavelengths only, because these
objects radiate most of their flux at long wavelengths
\citep{habing,bussorev}. They demonstrated that previous analysis
were underestimating the
IR contribution to the LFGCS.\\
In recent years, more precise parallaxes for Galactic
C-rich stars have become available \citep{vl07}.
Moreover, a new Period-Luminosity (hereafter P-L)
relation for C-rich Red Giants has been presented by
\citet{white} for the Large Magellanic Cloud (LMC). The
aim of this paper is to derive a new LFGCS, by re-analyzing the
C-rich sample of \citet{gua06} with the aforementioned upgrades
and new selection criteria. We also wish to compare the LFGCS with
theoretical AGB models \citep{cris11} and with a recent
observational Carbon Stars Luminosity Function derived for the
LMC \citep{gu12}. Possibly, this comparison could shed
light on the dependence of the Carbon Stars Luminosity
Function on metallicity and, thus, on the
hosting systems.

\begin{figure}[t]
   \centering
   {\includegraphics[width=8cm]{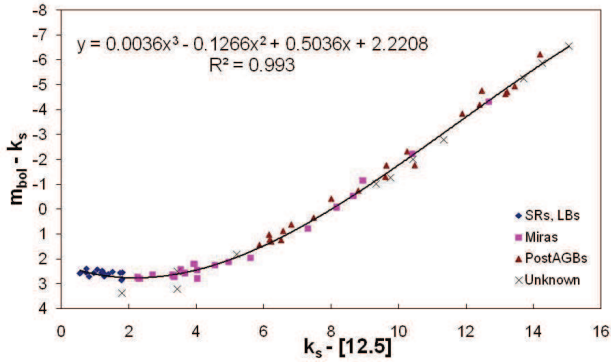}}
   \caption{Bolometric correction exploited to determine the apparent magnitudes. See text for details.}
   \label{fig:bc}
\end{figure}

In Section \ref{sec:sample} we describe the methods adopted to
build up our Carbon stars sample; updated Bolometric Corrections
are presented in Section \ref{sec:bc}; the new LFGCS is shown and discussed
in Section \ref{sec:lf}. Our conclusions follow in Section
\ref{sec:concl}.

\section{C-stars sample} \label{sec:sample}

In a previous paper, \citet{gua06} presented the observational
LFGCS by collecting a sample of already published Carbon-rich AGB
stars of the Milky Way. In that work, the sample was mainly made
of stars that had mass loss estimates and for which distance
estimates and/or mid-IR
photometry were available. \\
Distance estimates were mainly taken from (in order of
priority):
\begin{enumerate}
\item{\citealt{bg05} (who re-analysed data from the original Hipparcos release);} 
\item{\citealt{gro02};} 
\item{ \citealt{sol01} and \citealt{loup}}.
\end{enumerate}
%Moreover, for some single objects we adopted estimates from other
%sources in literature. \\
Near-IR data were retrieved from the ground-based 2MASS survey
\citep{cutri}. Mid-IR photometry was taken from space telescopes
ISO (SWS), MSX and IRAS (LRS). Selection criteria adopted to build
up the sample have been (in order of priority):
\begin{itemize}
\item{sources with ISO-SWS photometry;} 
\item{sources with enough
mid-IR photometric data to apply Bolometric Corrections;}
\item{application of the P-L relation for Mira variable sources
from various references that are shown in the Tables of \citet{gua06}.}
\end{itemize}
The sample presented in \citet{gua06} consists of 230 sources (see
their Fig. 8).

In recent years updated distance estimates for Galactic
stars have become available, thanks to a release of
revised Hipparcos astrometric data \citep{vl07}. Moreover, a new
P-L relation for C-stars has been published by \citet{white}. We
construct a new LFGCS starting from the sample of
\citet{gua06}, by using the aforementioned upgrades and by
following more stringent selection criteria, i.e.:
\begin{enumerate}
\item{We keep in the sample only sources with distances
derived from \citet{vl07}, selecting stars whose uncertainty
in parallaxes is lower than half the parallax value.
These sources must have ISO-SWS photometry or, on second option,
photometric observations at mid-IR wavelengths that allow the
application of the Bolometric Corrections (hereafter BC).
These sources belong to different variability classes
(i.e. Semiregulars and Miras).}
\item{If distances from
Hipparcos are not available, we include the sources for
which we can use the P-L relation presented by
\citet{white} to directly determine the bolometric magnitude
($M_{bol}$) of the star, even if this relation has been
derived for LMC sources. \citet{fea06} demonstrated that
both the slope and the zero-point of the P-L relation for Galactic
Carbon stars are similar to those derived for LMC Carbon stars.
We therefore decided to adopt the P-L relation for LMC
sources given in \citet{white} and thus we
follow their suggestion \citep{fea06}. This relation has been applied
to "bona-fide" Mira-type stars only, thus excluding from the
sample variable stars whose classification is uncertain. This
approach implies notable changes with respect to the method
adopted in \citet{gua06}, who determined $M_{bol}$ by coupling
their photometric analysis with distances determined from
Period-Luminosity relations presented in various works [references
shown in the Tables of \citet{gua06}]. However, the published P-L
relations were obtained by calculating the apparent magnitude
($m_{bol}$) without considering mid-IR data (wavelengths larger
than 8 $\mu$m), whose contribution for Carbon stars (and in
particular for Mira variables) cannot be neglected. On the other hand,
\citet{gua06} properly determined $m_{bol}$ analyzing the spectrum
at least up to 12 $\mu$m (MSX, IRAS-LRS) and, when available, up
to 45 $\mu$m (ISO-SWS). Thus, these distances could be
overestimated, leading to an overestimation of $M_{bol}$ (up to
half magnitude). Our procedure does not fix the need of
mid-IR data, but, with respect to \citet{gua06}, minimize the
uncertainties in the analysis).}
\item{We exclude from the sample sources that seem
to have a "reliable" estimate of the distance from \citet{vl07},
but have estimates of the absolute bolometric magnitude fainter
than $-3.5$. In fact, these stars cannot be AGB stars, but they
could be Giants belonging to a binary system polluted by an
already extinct AGB companion.} 
\item{Unlike \citet{gua06}, we
exclude from the sample CJ stars and C(R) stars (due to the
elusive nature of these objects; see e.g.
\citealt{abiaise,piersanti}).}
\item{We add to the sample
sources from \citet{white,lb05} not included
in \citet{gua06} that follow the previous
criteria.}
\end{enumerate}
The new sample consists of 102 sources, 32 stars
exploiting distances from \citealt{vl07} and the remaining
70 from the PL-relation method. There are 72 Miras and
30 of other variability types (see
the Online material for a list of the considered objects and their
properties). With respect to \citet{gua06}, the number of C-stars
is greatly reduced (more than halved), but we reckon that the
sample robustness is definitely improved. In fact, our
sample contains both dust enshrouded stars (i.e. stars with high
mass-loss rates, see \citealt{gua06}) and optically bright stars.

\begin{figure}[t]
   \centering
   {\includegraphics[width=8cm]{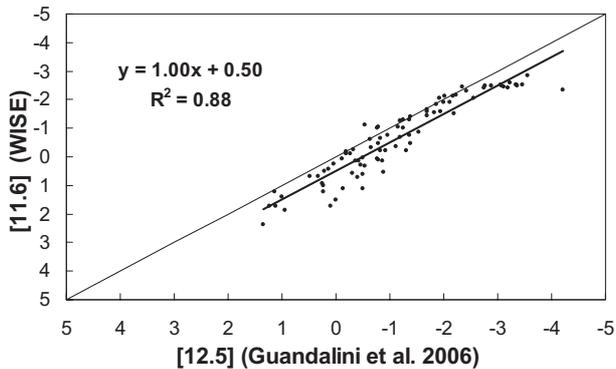}}
   \caption{Comparison between the apparent magnitudes obtained by WISE at 11.6 $\mu$m and
   the ones used by us at 12.5 $\mu$m for the sources in our sample.} 
   \label{fig:wise}
\end{figure}

We also verified if the two adopted methods offer us comparable
results. There are 5 Mira-type sources for which we have both a
reliable estimate of the distance from Hipparcos and an estimate
of the period of variability. Unfortunately, the
catalogues we use do not have mid-IR data for two of them
(\object{U Cyg} and \object{RZ Peg}). Notwithstanding, we
exploit the photometric data from the recently released WISE
catalogue \citep{cutri12}, in particular the fluxes for the filter
centered at 11.6 $\mu$m. Note that our analysis exploits a
different filter centered at 12.5 $\mu$m \citep[TIRCAM - see
][]{busso07}. We compare the fluxes obtained for the stars of our
sample in these two photometric filters by WISE and by the
catalogues used in this paper. In Fig.\ref{fig:wise} we observe a
good correlation between the two sets of observations, with the
ones in the 12.5 $\mu$m band on average 0.5 mag brighter than the
ones in the 11.6 $\mu$m band. Therefore, we convert the WISE data
at 11.6 $\mu$m according to the equation:
\begin{equation}
[11.6]_{WISE}=[12.5]_{G2006}+0.50
\end{equation}

Hereafter we report our analysis for the 5 stars for which
distance estimates from both the Hipparcos' catalogue and the P-L
relation are available:

\begin{enumerate}
  \item \object{R Lep}: bolometric magnitude obtained thanks to mid-IR data and Hipparcos' distance
  is $-$5.48, while if we take advantage of WISE data for mid-IR we obtain a bolometric
   magnitude of $-$5.55. The estimate obtained with the P-L method is $-$4.81 (around half magnitude fainter).
  \item  \object{S Cep}: bolometric magnitude obtained thanks to mid-IR data and Hipparcos' distance
  is $-$5.42, while if we exploit WISE data for mid-IR we obtain a bolometric
   magnitude of $-$5.37. The estimate obtained with the P-L method is $-$4.96 (around half magnitude fainter).
   \item  \object{U Cyg}: bolometric magnitude obtained thanks to WISE data for mid-IR data and
   Hipparcos distance is $-$5.40. The estimate obtained with the P-L method is $-$4.90 (around half magnitude fainter).
  \item \object{V Oph}:  bolometric magnitude obtained thanks to mid-IR data and Hipparcos' distance
  is $-$2.42, while if we use WISE data for mid-IR we obtain a bolometric
   magnitude of $-$2.42. The estimate obtained with the P-L method is $-$4.41.
  \item \object{RZ Peg}: bolometric magnitude obtained thanks to WISE data for mid-IR data and
   Hipparcos distance is $-$1.69. The estimate obtained with the P-L method is $-$4.84.
\end{enumerate}

The large differences in the last two sources (V Oph and
RZ Peg) could be due to unreliable estimates of the
distance from the Hipparcos' catalogue. For these stars
we adopt the estimate given by the P-L method, otherwise
their luminosity would be inconsistent with those
characterizing AGB stars.

The three remaining sources show that the
P-L method give bolometric magnitudes around half magnitude
fainter than the estimates obtained with mid-IR photometry and
Hipparcos distances.
The sample for which we can apply both methods is very
small, therefore this comparison cannot give us clear indications
about possible systematic biases between the two methods adopted
in this analysis. Moreover, we note that there may be some objects in
the sample with substantial errors in $M_{bol}$. \\
We found that the estimates of the absolute luminosity obtained
taking advantage of the WISE data are very similar to the ones
obtained adopting other mid-IR catalogues: in the study of the LF
we are going to use WISE mid-IR photometry for \object{R
Lep}, \object{S Cep} and \object{U Cyg}. The analysis of AGB stars of 
different chemical types with the WISE catalogue will be subject of a future paper.

\section{Bolometric Corrections} \label{sec:bc}

In order to properly determine the apparent magnitude of our
objects, we calculate again their BCs. In doing it, we consider a
larger sample with respect to the one determined to construct the
LFGCS. In particular, we analyze a large sample of intrinsic
Carbon stars observed by ISO-SWS, without assuming as a constraint
the distance estimate (note that the distance is not needed to
estimate the BC of a stellar object), in order to obtain their
$m_{bol}$. We adopt as a photometric index $m_{bol}-K_s$
(see Figure \ref{fig:bc}). Thus, as a by-product of our analysis,
we produce a new BC, which represent another important
improvement from to the ones presented in \citet{gua06} (see their
Fig. 5).The BC presents a smaller peak value than the one obtained in
\citet{gua06}, even if we followed the same procedure to derive it. This is probably a consequence
of using a different sample with respect to \citet{gua06}.

\section{The new Luminosity Function} \label{sec:lf}

\begin{figure}
   \centering
   {\includegraphics[width=8cm]{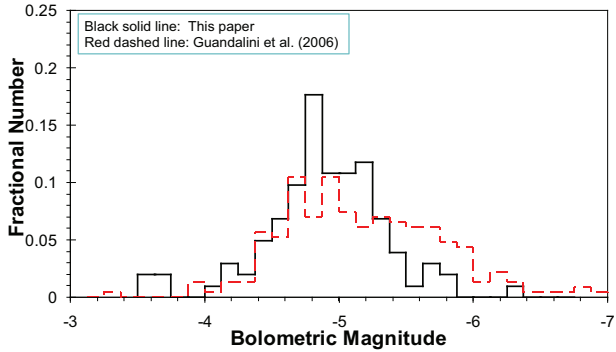}}
   \caption{The Observational Luminosity Function of Galactic Carbon Stars derived in this paper (black solid line) compared with the Observational Luminosity Function of Galactic Carbon Stars presented by \citealt{gua06} (red dashed line).}
   \label{fig:lfold}
\end{figure}

In Fig. \ref{fig:lfold} we show the LFGCS derived in this paper
(black solid line) and the one extracted from the C-stars sample
analyzed by \citealt{gua06} (red dashed line). The updated
observational LFGCS confirms the behavior of the previous one at
low and intermediate luminosities, with appreciable numbers
starting at $M_{bol}\sim -4$ and the peak placed around $M_{bol}=
-4.9$. The average uncertainty in the determination of
$M_{bol}$ is around $\pm0.3$ (see \citealt{white08}). The main
difference between the new and the old observational LFGCS
consists in the high luminosity tail. In fact, the new
LFGCS is truncated at $M_{bol}= -5.5$, and the high luminosity
tail practically disappears. The absence of the high luminosity
tail in the new LFGCS derives from the use of the aforementioned
recent data and from new selection criteria. Thus, we demonstrate
that the revised version of the Hipparcos catalogue \citep{vl07}
and a different choice in the exploitation of the P-L relations
lead to significant changes in the derived LFGCS.

\begin{figure}
   \centering
   {\includegraphics[width=8cm]{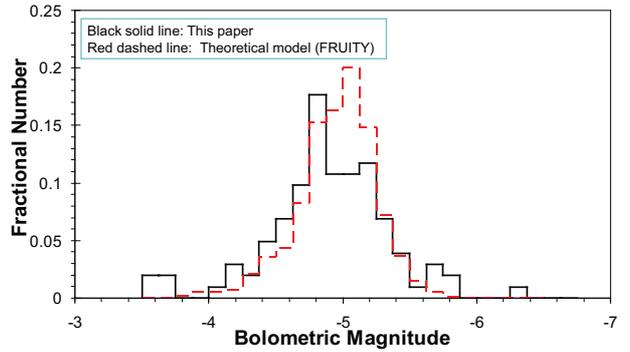}}
   \caption{The Observational Luminosity Function of Galactic Carbon Stars derived in this paper (black solid line),
   compared with the theoretical LFGCS published by \citealt{cris11} (red dashed line).}
   \label{fig:lfnew}
\end{figure}

The new LFGCS is in agreement with extant theoretical studies
\citep{cris11} as shown in Fig.\ref{fig:lfnew}. The
theoretical LFGCS has been constructed by evaluating the
contributions from all Galactic stars with different masses, ages
and metallicities currently evolving along the AGB. The lower and
upper mass on the AGB are estimated by interpolating the physical
inputs (main sequence lifetime, AGB lifetime, bolometric
magnitudes along the AGB) on the grid of computed models
(M$_{min}< $M/M$_\odot < 3.0$ M$_\odot$, with M$_{min}$ depending
on the metallicity). The theoretical LFGCS is nearly independent
from the assumed Star Formation Rate, Initial Mass Function and
metallicity distribution (see \citealt{cris11} for details). The
agreement between observational and theoretical LFGCS supports the
validity of the adopted stellar models, whose intrinsic
uncertainties (in particular the treatment of convection and the
mass-loss history) restrain their predictive power.

We remark, however, that major uncertainties still affect the
observational Luminosity Function of Galactic Carbon Stars. A
giant step toward a better comprehension of Galactic C-stars
and, thus, to the associated Luminosity Function, will be possible
with the data from the GAIA mission. In fact, GAIA will produce,
with an unprecedent precision, distance estimates for hundreds of
thousands of Galactic C-stars \citep{eyer}. Moreover, its
continuous sky mapping over the mission time (an average of 70
measurements per object is currently planned) will provide more
stringent constraints on the Period-Luminosity relations
characterizing Mira and Semi-Regular variable stars.
Moreover, we remind that the P-L relation was derived by
\citet{white} exploiting only observation at near-IR wavelengths:
a revision of the P-L relations considering also mid-IR photometry
is needed. This further improvement will be possible only when
dedicated surveys will release mid-IR data for Galactic and LMC
Miras\footnote{A good candidate could be the AMICA
infrared camera mounted on the IRAIT telescope.}.

\begin{figure}
   \centering
   {\includegraphics[width=8cm]{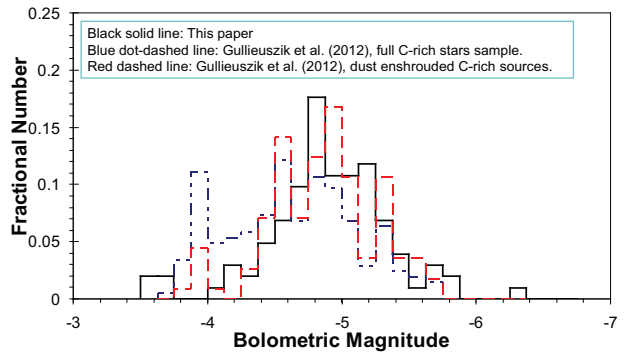}}
   \caption{The Observational Luminosity Function of Galactic Carbon Stars derived in this paper (black solid line),
   compared with the Luminosity Function of Carbon Stars in the Large Magellanic Cloud presented by \citealt{gu12} for the entire C-rich sample (blue dot-dashed line) and for the sub-sample of the dust-enshrouded sources (red dashed line).}
   \label{fig:lfgu}
\end{figure}

The problem of the distance determination does not affect the
Luminosity Function of Carbon Stars in the Large Magellanic Cloud,
since all C-stars belonging to that system can be considered at a
fixed distance with only moderate depth. Thus, any difference in
$m_{bol}$ implies a rescaled difference in luminosity. In
Figure \ref{fig:lfgu} we report the Luminosity Function of Carbon
stars in LMC derived by \citet{gu12}. We note that the
Luminosity Function, as obtained from their full sample
(dot-dashed curve), is more weighted toward fainter $M_{bol}$ than
our LFGCS. We should note, however, that the same authors claimed
a possible misclassification of faint C-rich stars. Moreover, the
sample presented by \citet{gu12} also contains C(J) stars, while
our LFGCS only contains C(N) stars, i.e. AGB stars currently
experiencing Third Dredge Up episodes. It is worth to remind that
the origin of C(J) stars is still unknown and that a considerable
percentage of these stars show infrared emission associated with
silicate dust (see e.g. \citealt{ruben}), while amorphous carbon
is expected to dominate the atmospheres of C(N) stars. Thus, we
suspect that C(J) stars could be classified as dust-free in the
sample of \citet{gu12}. Considering the aforementioned problems,
we reckon that the right sample to be compared with our LFGCS is
the dusty one presented by \citealt{gu12} (see their Figure 6).
For this reason, in Figure \ref{fig:lfgu} we also report this
sub-sample (dashed curve). The latter (see also \citealt{cohen})
peaks at the same Bolometric Magnitude and presents a shape very similar 
to the LFGCS presented in this paper.

This seems to suggest that, for intermediate and
large metallicities, the Luminosity Function of Carbon Stars
weakly depends on the initial metal content and that its magnitude
range is nearly the same ($-4.5<M_{bol}<-5.5$). The main
difference between the two Luminosity Functions shown in Fig.
\ref{fig:lfgu} is a very small shift to lower luminosities for the
sources of the Large Magellanic Cloud. This fact is expected from
theoretical calculations (see \citealt{cris11}, in particular
their Fig. 11). At lower metallicities, in fact, the reduced
oxygen content makes the carbon phase (C/O>1) on the AGB easier 
to be reached at lower luminosities (in earlier evolutionary stages).
Moreover, the enhanced TDU efficiency and the
larger contribution from lower masses (see \citealt{cri09})
further weight the Luminosity Function of Carbon Stars to lower
luminosities. New observations of Carbon Stars in low metallicity
environments, such as the Small Magellanic Cloud and Dwarf
galaxies (see \citealt{sloan} and references therein), could
further shed light on this problem. This analysis, however, is
beyond the scope of this paper.

\onltab{
\onecolumn
\begin{longtab}
\centering
\begin{longtable}{ccccc}
\caption{\label{table:1} Sample used to estimate the Luminosity Function.}\\
\hline\hline
Source & Var. Type & Distance (Kpc) & Period (days) & Absolute \\
name & (GCVS) & \citep{vl07}  & (GCVS) & Magnitude \\
\hline
\endfirsthead
\caption{continued.}\\
\hline\hline
Source & Var. Type & Distance (Kpc) & Period (days) & Absolute \\
name & (GCVS) & \citep{vl07}  & (GCVS) & Magnitude \\
\hline
\endhead
\hline
\endfoot
\object{W Cas} &   Mira  &   $-$ &  405.57   &   $-$4.75 \\
\object{HV Cas} &   Mira  &   $-$ &  527   &   $-$5.04 \\
\object{X Cas} &   Mira  &   $-$ &  422.84   &   $-$4.80 \\
\object{YY Tri} &   Mira  &   $-$ &  624   &   $-$5.23 \\
\object{R For} &   Mira  &   $-$ &  388.73   &   $-$4.71 \\
\object{V384} Per &   Mira  &   $-$ &  535   &   $-$5.06 \\
\object{V718} Tau &   Mira  &   $-$ &  405   &   $-$4.75 \\
\object{AU Aur} &   Mira  &   $-$ &  400.5   &   $-$4.74 \\
\object{R Ori} &   Mira  &   $-$ &  377.1   &   $-$4.67 \\
\object{R Lep} &   Mira  &   0.47 &  427.07   &   $-$5.55 \\
\object{NAME SHV F4488} &   Mira  &   $-$ &  573   &   $-$5.14 \\
\object{WBP 14} &   Mira  &   $-$ &  325  &   $-$4.51 \\
\object{V370 Aur} &   Mira  &   $-$ &  683   &   $-$5.33 \\
\object{QS Ori} &   Mira  &   $-$ &  476   &   $-$4.93 \\
\object{V617 Mon} &   Mira  &   $-$ &  375   &   $-$4.67 \\
\object{ZZ Gem} &   Mira  &   $-$ &  317   &   $-$4.48 \\
\object{V636 Mon} &   Mira  &   $-$ &  543   &   $-$5.08 \\
\object{V503 Mon} &   Mira  &   $-$ &  355   &   $-$4.61 \\
\object{RT Gem} &   Mira  &   $-$ &  350.4   &   $-$4.59 \\
\object{CG Mon} &   Mira  &   $-$ &  419.11   &   $-$4.79 \\
\object{CL Mon} &   Mira  &   $-$ &  497.15   &   $-$4.98 \\
\object{R Vol} &   Mira  &   $-$ &  453.6   &   $-$4.88 \\
\object{HX CMa} &   Mira  &   $-$ &  725   &   $-$5.40 \\
\object{VX Gem} &   Mira  &   $-$ &  379.4   &   $-$4.68 \\
\object{V831 Mon} &   Mira  &   $-$ &  319   &   $-$4.49 \\
\object{V346 Pup} &   Mira  &   $-$ &  571   &   $-$5.13 \\
\object{FF Pup} &   Mira  &   $-$ &  436   &   $-$4.83 \\
\object{IQ Hya} &   Mira  &   $-$ &  397   &   $-$4.73 \\
\object{CQ Pyx} &   Mira  &   $-$ &  659   &   $-$5.29 \\
\object{CW Leo} &   Mira  &   $-$ &  630   &   $-$5.24 \\
\object{CZ Hya} &   Mira  &   $-$ &  442   &   $-$4.85 \\
\object{TU Car} &   Mira  &   $-$ &  258   &   $-$4.26 \\
\object{FU Car} &   Mira  &   $-$ &  365   &   $-$4.64 \\
\object{V354 Cen} &   Mira  &   $-$ &  150.4   &   $-$3.66 \\
\object{BH Cru} &   Mira  &   $-$ &  421   &   $-$4.80 \\
\object{V1132 Cen} &   Mira  &   $-$ &  560   &   $-$5.11 \\
\object{V Cru} &   Mira  &   $-$ &  376.5   &   $-$4.67 \\
\object{TT Cen} &   Mira  &   $-$ &  462   &   $-$4.90 \\
\object{RV Cen} &   Mira  &   $-$ &  446   &   $-$4.86 \\
\object{II Lup} &   Mira  &   $-$ &  580   &   $-$5.15 \\
\object{V CrB} &   Mira  &   $-$ &  357.63   &   $-$4.62 \\
\object{NP Her} &   Mira  &   $-$ &  448   &   $-$4.86 \\
\object{V Oph} &   Mira  &   0.24 &  297.21   &   $-$4.41 \\
\object{V2548 Oph} &   Mira  &   $-$ &  747   &   $-$5.43 \\
\object{V617 Sco} &   Mira  &   $-$ &  523.6   &   $-$5.04 \\
\object{V833 Her} &   Mira  &   $-$ &  540   &   $-$5.07 \\
\object{T Dra} &   Mira  &   $-$ &  421.62   &   $-$4.80 \\
\object{V1280 Sgr} &   Mira  &   $-$ &  523   &   $-$5.03 \\
\object{V5104 Sgr} &   Mira  &   $-$ &  655   &   $-$5.28 \\
\object{V1076 Her} &   Mira  &   $-$ &  609   &   $-$5.20 \\
\object{V627 Oph} &   Mira  &   $-$ &  452   &   $-$4.87 \\
\object{V821 Her} &   Mira  &   $-$ &  511   &   $-$5.01 \\
\object{V1417 Aql} &   Mira  &   $-$ &  617   &   $-$5.22 \\
\object{V874 Aql} &   Mira  &   $-$ &  145  &   $-$3.62 \\
\object{V2045 Sgr} &   Mira  &   $-$ &  451   &   $-$4.87 \\
\object{AI Sct} &   Mira  &   $-$ &  408   &   $-$4.76 \\
\object{V1420 Aql} &   Mira  &   $-$ &  676   &   $-$5.32 \\
\object{V1965 Cyg} &   Mira  &   $-$ &  625   &   $-$5.23 \\
\object{KL Cyg} &   Mira  &   $-$ &  526   &   $-$5.04 \\
\object{R Cap} &   Mira  &   $-$ &  345.13   &   $-$4.58 \\
\object{U Cyg} &   Mira  &   0.52 &  463.24   &   $-$5.40 \\
\object{BD Vul} &   Mira  &   $-$ &  430   &   $-$4.82 \\
\object{V Cyg} &   Mira  &   $-$ &  421.27   &   $-$4.80 \\
\object{V442 Vul} &   Mira  &   $-$ &  661   &   $-$5.29 \\
\object{RV Aqr} &   Mira  &   $-$ &  453   &   $-$4.88 \\
\object{V1426 Cyg} &   Mira  &   $-$ &  470   &   $-$4.92 \\
\object{S Cep} &   Mira  &   0.41 &  486.84   &   $-$5.37 \\
\object{V1568 Cyg} &   Mira  &   $-$ &  495   &   $-$4.97 \\
\object{RZ Peg} &   Mira  &   0.21 &  438.7   &   $-$4.84 \\
\object{LL Peg} &   Mira  &   $-$ &  696   &   $-$5.35 \\
\object{IZ Peg} &   Mira  &   $-$ &  486   &   $-$4.95 \\
\object{LP And} &   Mira  &   $-$ &  614   &   $-$5.21 \\
\hline
\object{VX And} &   SRA  &   0.39 &  375   &   $-$4.16 \\
\object{Z Psc} &   SRB  &   0.38 &  144   &   $-$4.40 \\
\object{R Scl} &   SRB  &   0.27 &  370   &   $-$3.71 \\
\object{TW Hor} &   SRB  &   0.32 &  158   &   $-$4.62 \\
\object{TT Tau} &   SRB  &   0.36 &  166.5   &   $-$4.24 \\
\object{W Ori} &   SRB  &   0.38 &  212   &   $-$5.76 \\
\object{W Pic} &   LB  &   0.78 &  $-$   &   $-$5.73 \\
\object{Y Tau} &   SRB  &   0.36 &  241.5   &   $-$4.70 \\
\object{NP Pup} &   LB  &   0.50 &  $-$   &   $-$5.02 \\
\object{X Cnc} &   SRB  &   0.34 &  195   &   $-$4.96 \\
\object{Y Hya} &   SRB  &   0.39 &  302.8   &   $-$4.76 \\
\object{X Vel} &   SR  &   0.36 &  140   &   $-$4.51 \\
\object{U Ant} &   LB  &   0.27 &  $-$   &   $-$5.22 \\
\object{VY UMa} &   LB  &   0.38 &  $-$   &   $-$4.75 \\
\object{V996 Cen} &   LB  &   0.64 &  $-$   &   $-$5.48 \\
\object{X TrA} &   LB  &   0.36 &  $-$   &   $-$5.74 \\
\object{V Pav} &   SRB  &   0.37 &  225.4   &   $-$4.91 \\
\object{S Sct} &   SRB  &   0.39 &  148   &   $-$4.73 \\
\object{V Aql} &   SRB  &   0.36 &  353   &   $-$5.19 \\
\object{UX Dra} &   SRA  &   0.39 &  168   &   $-$4.90 \\
\object{AQ Sgr} &   SRB  &   0.33 &  199.6   &   $-$4.11 \\
\object{TT Cyg} &   SRB  &   0.56 &  118   &   $-$4.22 \\
\object{AX Cyg} &   LB  &   0.37 &  $-$   &   $-$3.60 \\
\object{T Ind} &   SRB  &   0.58 &  320   &   $-$5.86 \\
\object{Y Pav} &   SRB  &   0.40 &  233.3   &   $-$5.11 \\
\object{V460 Cyg} &   SRB  &   0.62 &  180   &   $-$6.28 \\
\object{TX Psc} &   LB  &   0.28 &  $-$   &   $-$5.15 \\
\object{UY Cen} &   SR  &   0.67 &  114.6   &   $-$5.73 \\
\object{RX Sct} &   LB  &   0.43 &  $-$   &   $-$4.29 \\
\object{RS Cyg} &   SRA  &   0.47 &  417.39   &   $-$4.39 \\
\end{longtable}
\end{longtab}
\twocolumn
}

\section{Conclusions} \label{sec:concl}

In this paper we present a revised observational LFGCS. New
available data (revised Hipparcos distances from \citealt{vl07}
and Period-Luminosity relations from \citealt{white}) and more
stringent criteria have been adopted to select the C-star sample.
We confirm the LFGCS of \citet{gua06} at low and intermediate
luminosities. On the contrary, at high luminosities the new LFGCS
is abruptly truncated at $M_{bol}= -5.5$, in agreement with extant
theoretical studies \citep{cris11}. The disappearance of the high
luminosity tail derives from the use of the aforementioned recent
data and new selection criteria.

The Luminosity Functions of dusty Carbon Stars in the
Large Magellanic Cloud (see \citealt{gu12} for a recent
derivation) peaks at the same Bolometric Magnitude with respect to
its Galactic counterpart, presenting a very similar shape. This
seems to suggest that the Luminosity Functions of Carbon Stars
slightly depends on the initial metal content and that its
magnitude range is nearly the same ($-4.5<M_{bol}<-5.5$) in
stellar systems with intermediate and large metallicities, as
suggested by \citet{cris11}. Observations are well reproduced by
AGB theoretical models. This demonstrates the goodness in the
choice of critical parameters affecting their evolution, such as
the treatment of convection or the mass-loss history, whose
complex interplay determines the duration and the luminosity
of the C-rich phase.

\begin{acknowledgements}
The authors deeply thank Oscar Straniero and Maurizio Busso for enlightening
discussions on stellar evolution and for a careful reading of this
manuscript. The authors deeply thank the referee for an extensive and helpful
review, containing very relevant scientific advice. 
S.C. acknowledges financial support from the FIRB2008
program (RBFR08549F-002) and from the PRIN-INAF 2011 project
"Multiple populations in Globular Clusters: their role in the
Galaxy assembly".
This publication makes use
VizieR catalogue access tool, CDS, Strasbourg, France and of data products
from the Wide-field Infrared Survey Explorer, which is a joint project
of the University of California, Los Angeles, and the Jet Propulsion Laboratory/California Institute of Technology, funded by the National Aeronautics and Space Administration.
\end{acknowledgements}

\bibliographystyle{aa}
\bibliography{LFacc}

\begin{thebibliography}{27}
\expandafter\ifx\csname natexlab\endcsname\relax\def\natexlab#1{#1}\fi

\bibitem[{{Abia} \& {Isern}(2000)}]{abiaise}
{Abia}, C. \& {Isern}, J. 2000, \apj, 536, 438

\bibitem[{{Bergeat} \& {Chevallier}(2005)}]{bg05}
{Bergeat}, J. \& {Chevallier}, L. 2005, \aap, 429, 235

\bibitem[{{Busso} {et~al.}(1999){Busso}, {Gallino}, \& {Wasserburg}}]{bussorev}
{Busso}, M., {Gallino}, R., \& {Wasserburg}, G.~J. 1999, \araa, 37, 239

\bibitem[{{Busso} {et~al.}(2007){Busso}, {Guandalini}, {Persi}, {Corcione}, \&
  {Ferrari-Toniolo}}]{busso07}
{Busso}, M., {Guandalini}, R., {Persi}, P., {Corcione}, L., \&
  {Ferrari-Toniolo}, M. 2007, \aj, 133, 2310

\bibitem[{{Cohen} {et~al.}(1981){Cohen}, {Persson}, {Elias}, \&
  {Frogel}}]{cohen}
{Cohen}, J.~G., {Persson}, S.~E., {Elias}, J.~H., \& {Frogel}, J.~A. 1981,
  \apj, 249, 481

\bibitem[{{Cristallo} {et~al.}(2011){Cristallo}, {Piersanti}, {Straniero},
  {Gallino}, {Dom{\'{\i}}nguez}, {Abia}, {Di Rico}, {Quintini}, \&
  {Bisterzo}}]{cris11}
{Cristallo}, S., {Piersanti}, L., {Straniero}, O., {et~al.} 2011, \apjs, 197,
  17

\bibitem[{{Cristallo} {et~al.}(2009){Cristallo}, {Straniero}, {Gallino},
  {Piersanti}, {Dom{\'{\i}}nguez}, \& {Lederer}}]{cri09}
{Cristallo}, S., {Straniero}, O., {Gallino}, R., {et~al.} 2009, \apj, 696, 797

\bibitem[{{Cutri} {et~al.}(2003){Cutri}, {Skrutskie}, {van Dyk}, {Beichman},
  {Carpenter}, {Chester}, {Cambresy}, {Evans}, {Fowler}, {Gizis}, {Howard},
  {Huchra}, {Jarrett}, {Kopan}, {Kirkpatrick}, {Light}, {Marsh}, {McCallon},
  {Schneider}, {Stiening}, {Sykes}, {Weinberg}, {Wheaton}, {Wheelock}, \&
  {Zacarias}}]{cutri}
{Cutri}, R.~M., {Skrutskie}, M.~F., {van Dyk}, S., {et~al.} 2003, {2MASS All
  Sky Catalog of point sources.}

\bibitem[{{Cutri} {et~al.}(2012){Cutri}, {Wright}, {Conrow}, {Bauer},
  {Benford}, {Brandenburg}, {Dailey}, {Eisenhardt}, {Evans}, {Fajardo-Acosta},
  {Fowler}, {Gelino}, {Grillmair}, {Harbut}, {Hoffman}, {Jarrett},
  {Kirkpatrick}, {Leisawitz}, {Liu}, {Mainzer}, {Marsh}, {Masci}, {McCallon},
  {Padgett}, {Ressler}, {Royer}, {Skrutskie}, {Stanford}, {Wyatt}, {Tholen},
  {Tsai}, {Wachter}, {Wheelock}, {Yan}, {Alles}, {Beck}, {Grav}, {Masiero},
  {McCollum}, {McGehee}, {Papin}, \& {Wittman}}]{cutri12}
{Cutri}, R.~M., {Wright}, E.~L., {Conrow}, T., {et~al.} 2012, {Explanatory
  Supplement to the WISE All-Sky Data Release Products}, Tech. rep.

\bibitem[{{Eyer} {et~al.}(2012){Eyer}, {Palaversa}, {Mowlavi}, {Dubath},
  {Anderson}, {Evans}, {Lebzelter}, {Ripepi}, {Szabados}, {Leccia}, \&
  {Clementini}}]{eyer}
{Eyer}, L., {Palaversa}, L., {Mowlavi}, N., {et~al.} 2012, \apss, 49

\bibitem[{{Feast} {et~al.}(2006){Feast}, {Whitelock}, \& {Menzies}}]{fea06}
{Feast}, M.~W., {Whitelock}, P.~A., \& {Menzies}, J.~W. 2006, \mnras, 369, 791

\bibitem[{{Groenewegen} {et~al.}(2002){Groenewegen}, {Sevenster}, {Spoon}, \&
  {P{\'e}rez}}]{gro02}
{Groenewegen}, M.~A.~T., {Sevenster}, M., {Spoon}, H.~W.~W., \& {P{\'e}rez}, I.
  2002, \aap, 390, 511

\bibitem[{{Guandalini} {et~al.}(2006){Guandalini}, {Busso}, {Ciprini},
  {Silvestro}, \& {Persi}}]{gua06}
{Guandalini}, R., {Busso}, M., {Ciprini}, S., {Silvestro}, G., \& {Persi}, P.
  2006, \aap, 445, 1069

\bibitem[{{Gullieuszik} {et~al.}(2012){Gullieuszik}, {Groenewegen}, {Cioni},
  {de Grijs}, {van Loon}, {Girardi}, {Ivanov}, {Oliveira}, {Emerson}, \&
  {Guandalini}}]{gu12}
{Gullieuszik}, M., {Groenewegen}, M.~A.~T., {Cioni}, M.-R.~L., {et~al.} 2012,
  \aap, 537, A105

\bibitem[{{Habing}(1996)}]{habing}
{Habing}, H.~J. 1996, \aapr, 7, 97

\bibitem[{{Hedrosa} {et~al.}(2013){Hedrosa}, {Abia}, {Busso}, {Cristallo},
  {Dom{\'{\i}}nguez}, {Palmerini}, {Plez}, \& {Straniero}}]{ruben}
{Hedrosa}, R., {Abia}, C., {Busso}, M., {et~al.} 2013, ArXiv e-prints

\bibitem[{{Iben}(1981)}]{iben}
{Iben}, Jr., I. 1981, \apj, 246, 278

\bibitem[{{Izzard} {et~al.}(2004){Izzard}, {Tout}, {Karakas}, \& {Pols}}]{izza}
{Izzard}, R.~G., {Tout}, C.~A., {Karakas}, A.~I., \& {Pols}, O.~R. 2004,
  \mnras, 350, 407

\bibitem[{{Le Bertre} {et~al.}(2005){Le Bertre}, {Tanaka}, {Yamamura},
  {Murakami}, \& {MacConnell}}]{lb05}
{Le Bertre}, T., {Tanaka}, M., {Yamamura}, I., {Murakami}, H., \& {MacConnell},
  D.~J. 2005, \pasp, 117, 199

\bibitem[{{Loup} {et~al.}(1993){Loup}, {Forveille}, {Omont}, \& {Paul}}]{loup}
{Loup}, C., {Forveille}, T., {Omont}, A., \& {Paul}, J.~F. 1993, \aaps, 99, 291

\bibitem[{{Piersanti} {et~al.}(2010){Piersanti}, {Cabez{\'o}n}, {Zamora},
  {Dom{\'{\i}}nguez}, {Garc{\'{\i}}a-Senz}, {Abia}, \& {Straniero}}]{piersanti}
{Piersanti}, L., {Cabez{\'o}n}, R.~M., {Zamora}, O., {et~al.} 2010, \aap, 522,
  A80

\bibitem[{{Sch{\"o}ier} \& {Olofsson}(2001)}]{sol01}
{Sch{\"o}ier}, F.~L. \& {Olofsson}, H. 2001, \aap, 368, 969

\bibitem[{{Sloan} {et~al.}(2012){Sloan}, {Matsuura}, {Lagadec}, {van Loon},
  {Kraemer}, {McDonald}, {Groenewegen}, {Wood}, {Bernard-Salas}, \&
  {Zijlstra}}]{sloan}
{Sloan}, G.~C., {Matsuura}, M., {Lagadec}, E., {et~al.} 2012, \apj, 752, 140

\bibitem[{{Straniero} {et~al.}(2006){Straniero}, {Gallino}, \&
  {Cristallo}}]{stra06}
{Straniero}, O., {Gallino}, R., \& {Cristallo}, S. 2006, Nuclear Physics A,
  777, 311

\bibitem[{{van Leeuwen}(2007)}]{vl07}
{van Leeuwen}, F. 2007, \aap, 474, 653

\bibitem[{{Whitelock} {et~al.}(2006){Whitelock}, {Feast}, {Marang}, \&
  {Groenewegen}}]{white}
{Whitelock}, P.~A., {Feast}, M.~W., {Marang}, F., \& {Groenewegen}, M.~A.~T.
  2006, \mnras, 369, 751

\bibitem[{{Whitelock} {et~al.}(2008){Whitelock}, {Feast}, \& {van
  Leeuwen}}]{white08}
{Whitelock}, P.~A., {Feast}, M.~W., \& {van Leeuwen}, F. 2008, \mnras, 386, 313

\end{thebibliography}

\Online

\end{document}